# Adaptive Kinetic-Fluid Models for Expanding Plasmas

**Robert Arslanbekov** [a] **and Vladimir Kolobov** [a,b,1]

[a] CFD Research Corporation, Huntsville, AL 35806, USA
[b] The University of Alabama in Huntsville, AL 35899, USA

E-mail: vladimir.kolobov@cfdrc.com

**Abstract**. We describe gas dynamic and plasma effects for transient plasma jets under different conditions. Our Unified Flow Solver (UFS) is evaluated for simulations of mixed continuum-rarefied flows, transient gas expansion dynamics and collisionless plasma expansion. We identify further advances that are required for hybrid fluid-kinetic simulations of laser-ablation and EEE-induced plasma jets.

## 1. Introduction

Expanding plasmas appear in various fields of space science and industrial technologies. Two types of expanding plasmas can be distinguished. The first type includes dense plasma expanding into vacuum or an ambient gas. These types of problems are characterized by a slow time scale, which is associated with adiabatic electron motion and by ion acceleration in plasma-induced electric fields. The second type includes plasmas formed inside gas plumes during propagation of fast ionization waves. These plasma jets are characterized by a fast time scale associated with ionization processes induced by electrons in a neutral gas; the ion motion at this time scale can be neglected. In particular, such plasma jets have been produced in atmospheric air inside plumes of rare gases, and are currently being studied for numerous applications [1]. The disparity of time scales for electron and ion transport allows substantial simplifications in modelling both types of plasma jets.

Here, we focus our attention on the first type of jets. An expansion of evaporated solid/liquid material into vacuum or an ambient background gas has been widely studied for pulsed laser ablation [2]. During the gas plume expansion, the evaporated material is accelerated as its initial internal energy is converted into kinetic energy. The resulting expansion velocity contains information about the properties of the hot dense gas before expansion including its ionization degree.

Micro-explosions also result in the formation of plasma jets expanding from a solid surface. During a jet expansion, dense non-ideal plasma turns into ideal plasma as the plasma density gradually decreases and inter-particle collisions are reduced. Plasma flares expanding into vacuum from arc-induced micro-explosions (ectons) on the cathode surface have been previously analyzed using both fluid and kinetic models [3,4,5]. The importance of hybrid fluid-kinetic approach to simulations of Explosive Electron Emission (EEE) and associated plasma expansion has been emphasized in Ref. [4]. However, no analysis of ambient gas effects on the dynamics of EEE-induced plasma jets has been reported so far.

In this paper, we describe simulations of gas dynamic and plasma effects for transient jets. We apply our Unified Flow Solver (UFS) [6] for mixed continuum-rarefied flows, investigate gas plume dynamics and collisionless plasma expansion. We identify further advances that are required for hybrid fluid-kinetic simulations of laser-ablation and EEE-induced plasma jets.

---
[1] To whom any correspondence should be addressed.

## 2. Simulations of Gas Plume Dynamics

Plume expansion produced by pulsed laser ablation in vacuum has been previously investigated experimentally and computationally by Direct Simulation Monte Carlo (DSMC) [7] for low-fluency regimes, when the plume ionization and plasma effects are negligible. DSMC simulations have demonstrated that hyperthermal kinetic energies, which are observed for ablated metals and graphite, can be explained through purely gas-dynamic considerations. The increase of particle kinetic energy is due to two effects: hydrodynamic plume acceleration into the forward direction and kinetic selection of fast particles in the on-axis region. The problem is characterized by two dimensionless parameters: $b = R/(v_T \tau)$ and $Kn = \lambda_0 / R$. The first one is the ratio of the spot radius $R$ to the distance travelled by the evaporated particles with thermal velocity $v_T$ during the duration of the laser pulse $\tau$. The second parameter is the Knudsen number, i.e. ratio of the particle mean free path to the spot radius, which defines initial collisionality degree of the plume [8]. For plume expansion into a steady ambient gas, additional parameters are the density ratio of the background particles and evaporated particles, $\eta = n_b / n_0$, and the temperature ratio, $\eta_T = T_b / T_0$.

We have first performed simulations of plume expansion into same background gas for different values of $n_0 / n_b$ using Boltzmann and Euler solvers in UFS. For simplicity, we used 2D-2V Boltzmann solver with the simplest BGK model for collisions. The base conditions, $\tau \sim 10$ ns, $R \sim 10$ μm, $v_T \sim 10^3$ m/s, correspond to $b \sim 1$. We used dimensionless variables in our calculations with velocity normalized to $v_T$, and density normalized to $n_0$. The computational domain consists of a unit box: $0 < x < 1$, $-0.5 < y < 0.5$. The boundary condition at the surface ($x = 0$), corresponds to an evaporating spot with dimensionless radius $R = 0.05$. Inside the spot, the dimensionless density, velocity and temperature are assumed to be equal to one during the power pulse, and then drop instantly to the ambient gas values, $\eta$ and $\eta_T$. The temperature ratio $\eta_T = 0.1$ was used in all simulations reported below.

Figure 1 shows an example of instantaneous spatial distributions of gas density, mean velocity and temperature at $t/\tau = 1.2$ for $\eta = 0.01$ and $Kn = 0.2$. The mesh was dynamically adapted based on gradients of gas density and temperature. The velocity mesh for the Boltzmann solver was selected as $-6 < \xi_x, \xi_y < 6$, with 24 cells in each velocity direction.

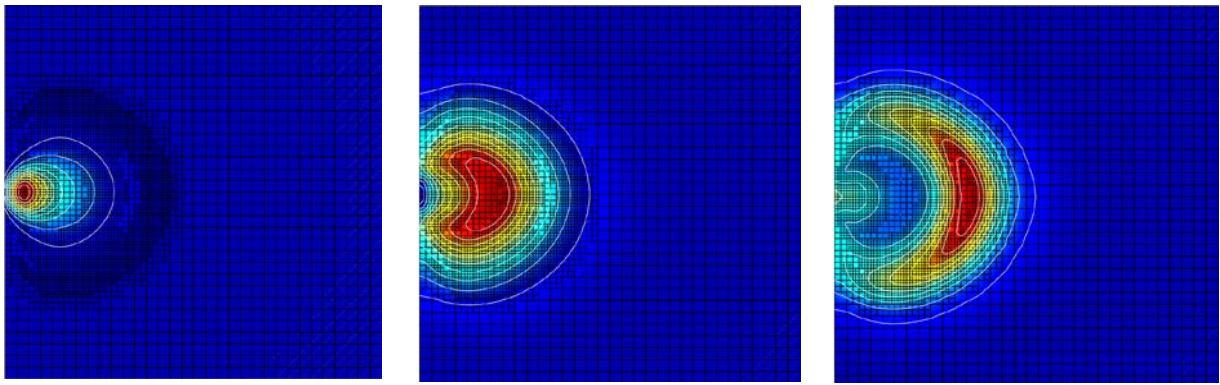

*Figure 1. Dynamically adapted mesh, gas density (left), mean velocity (center) and temperature (right) at $t/\tau = 1.2$ for density ratio $\eta = 0.01$ and $Kn = 0.2$.*

Simulations have been performed for different $\eta$ and $Kn$. Figure 2 shows a time dependence of the gas density on the axis at $L = 10R$ for $\eta = 0.1$, 0.01, and 0.001. Solid lines correspond to $Kn = 0.2$, dashed line is for *Kn* = 0.01. The calculated density variations correspond to the time-of-flight (TOF) distributions measured by density-sensitive detectors in experiments. It is seen that the shock wave propagates faster at lower background densities (lower $\eta$ ), and a decrease of *Kn* number results in a sharper front of the TOF function.

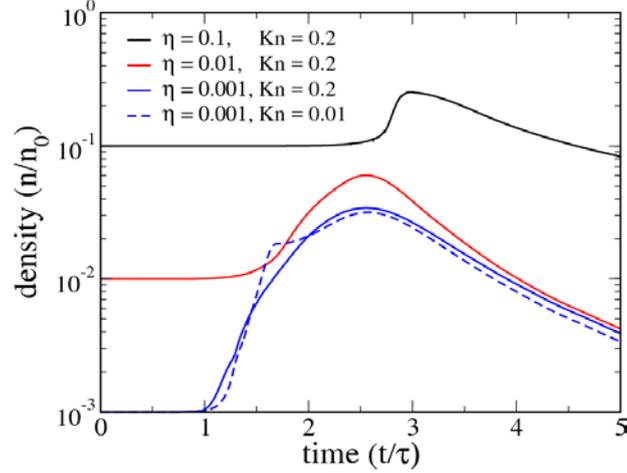

*Figure 2. Time of flight function for different initial density ratios at L=10R.*

## 3. Expansion of Evaporated Material into Vacuum or into Background Gas

*3.1. 1D Simulations of Cu Expansion into Argon with a Navier-Stokes Solver*
Expansion of evaporated metal plumes into vacuum or a background gas has been studied in relation to laser ablation for thin film deposition (see, e.g., Ref. [9,10] and references cited therein). We have first solved this problem with a multispecies Navier-Stokes (NS) solver. Following Ref. [10], we first considered a 1D expansion problem with a background Ar gas pressure of 200 mTorr and a temperature of 300 K at rest. The mean free path, $\lambda$, for Ar atoms under these initial is ~10 μm. At the beginning, we created a region of high density and high pressure of Cu atoms with density of $10^{19}$ cm$^{-3}$ and temperature 7000 K. The density ratio is thus ~1500, and the pressure ratio ~35000. The size of this initial Cu core is 50 μm (~5 initial Ar atom mean free paths and ~7000 Cu atom mean free paths). The computation domain of 15 cm (~15000 initial Ar atom mean free paths) was utilized. Because of such a large variation of the problem spatial scales, as well as moving strong shocks and discontinuities, a dynamically adapted grid was used with adaptation on species mass fractions, density, pressure and Mach number (each with a different threshold value).

The flow field profiles at two time instances are shown in Figure 3. One can see that shortly after the expansion starts, a density peak is formed due to gas compression (the snowplowing effect, see, e.g., Ref.[10]). This compression takes place at the Cu plume front. Further downstream from this front, there develops a moving shock propagating with a velocity of ~3000 m/s. At this time, the Cu plume expands over ~5 mm (or ~500 initial mean free paths of Ar atoms). At a later time of 108 μs, the shock expands over 12 cm, while the Cu plume decelerates to much smaller speeds, hardly moving in the preceding 100 μs. The shock propagates into the background gas and the density drops by 3 orders of magnitude. By comparing these results with those presented in Ref.[10] for a 1D plume expansion we observed close agreement considering that different metal and background species were simulated and a laser source was used (see below). In particular, at the expansion time of 100 μs, a negative flow velocity of ~200 m/s was predicted in Ref. [10] (which is an indication of a counter flow flux of Ar atoms), while NS predicts a negative velocity of 190 m/s at this time, see Figure 3, right.

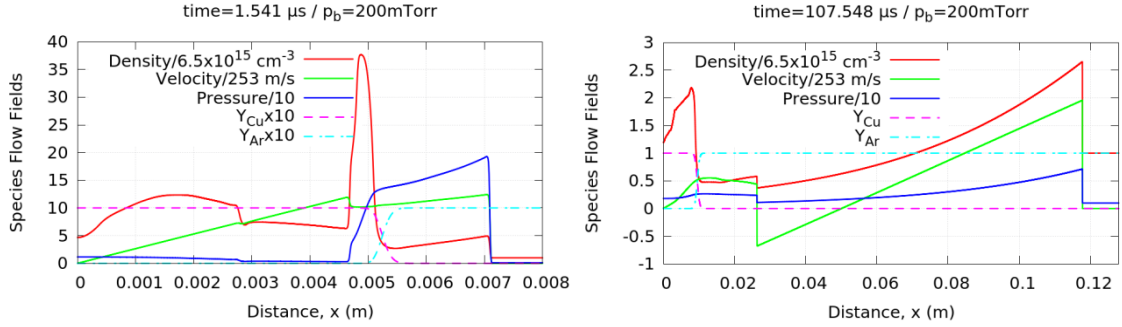

*Figure 3. Species flow fields obtained by NS solver at times of ~1.5 μs (left) and ~107 μs (right).*

It is well known that the presence of background gas radically changes the dynamics of metal plume expansion [10]. To study this effect, we have also conducted simulations for the background pressures of 20 mTorr and 2 Torr. The time evolutions of the peak velocities and temperatures are compared in Figure 4 for three background gas pressures. One can see that the plume propagation peak velocity and temperature decrease with increasing gas pressure, as has been previously observed in numerical and experimental studies. In addition to the free expansion simulations, we have carried out simulations by assuming source terms for mass density and energy, which are present during the action of a short laser pulse according to Ref. [10]. To simplify the analysis, we approximated the initial flow solutions from Ref. [9] obtained with source terms and carried out the subsequent computations. As one can see in Figure 4, the plume velocities of ~5000 m/s are predicted, while the previous free expansion simulations with the step-function initial conditions show smaller velocities by a factor of ~2. These results are in very good agreement with Ref. [10], where also a factor of 2 differences have been predicted. Also, we note the NS solver correctly reproduces typical velocities of the initial metal plume propagation, which are of the order of $10^4$ m/s, as has been observed experimentally and numerically, see Ref. [9]. One can also see that the predicted peak temperatures with the sources (reaching ~25000 K) are about a factor of 2 higher compared to the free expansion results. These findings emphasize the importance of metal content initial conditions for accurate predictions of plume expansion into vacuum or low background pressure environment.

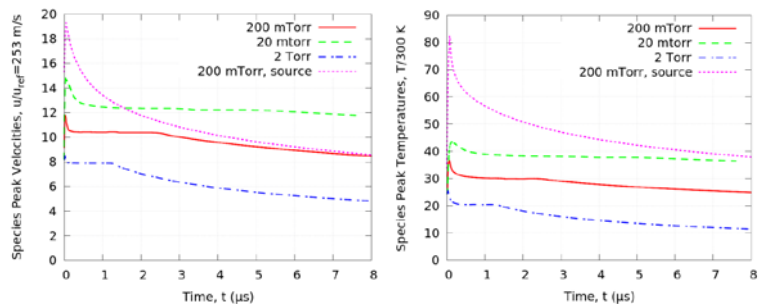

*Figure 4. Temporal evolutions of species mean velocities and temperatures for 20 mTorr, 200 mTorr and 2 Torr for the free expansion case. Also shown are evolutions in case with a laser source.*

### 3.2. 1D Simulations of Cu Plume Expansion into Argon with a Boltzmann Solver

We next study the problem of high-density Cu plume expansion into a low-pressure Ar gas by the multispecies Boltzmann kinetic solver. The same initial conditions are used as in the previously described simulations with the NS solver. The results for Ar pressure of 200 mTorr are shown in Figure 5 at different time instances. One can see that the velocity of the Cu plume front quickly increases to high values of ~16 × $u_{ref}$ (= 16×253 m/s) and compression of the background gas starts to occur as the Ar density in the vicinity of the Cu plume front increases several with respect to the background gas density. This corresponds to the snow-plow mechanism discussed in the literature.[9,10] At later times of ~1 μs, a compression wave in Ar gas develops forming a supersonic shock wave propagating over the background gas. The most interesting feature of the kinetic simulations is the fact that over the initial time of about 1–2 μs the Cu atom mean velocity and temperature differ significantly from those of the Ar gas. Indeed, Cu atoms attain velocities almost double of that of the Ar atoms, while Cu atoms temperatures are more than double of temperature of Ar atoms, reaching values of ~60 × $T_{ref}$ (= 300 K), or ~18000 K. At later times (t >1–2 μs), the macro-parameters of Ar and Cu species start to coincide with each other. At these times, energetic Cu atoms have travelled a distance of over several 100's of mean free paths and become in equilibrium with the background gas.

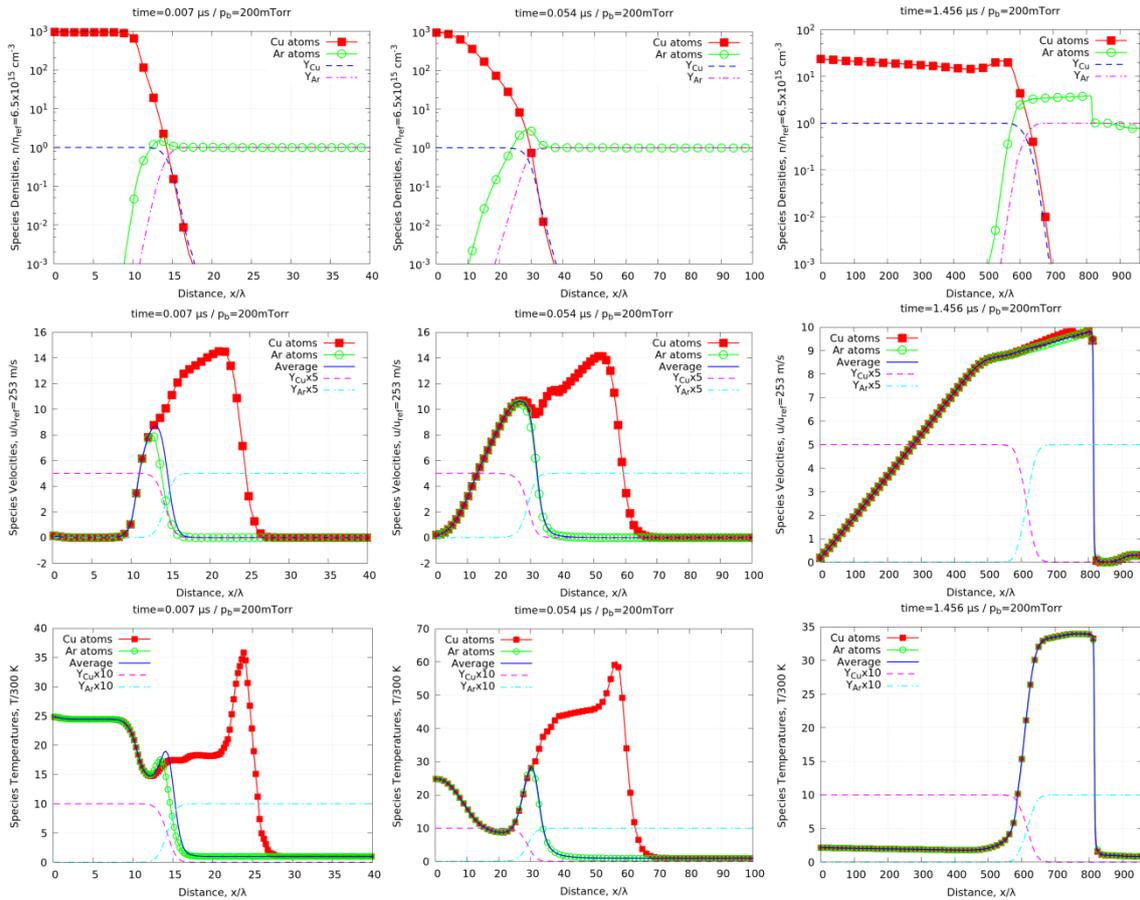

*Figure 5. Species macroparameters obtained by Boltzmann solver for 200 mTorr. Shown are species densities (top row), mean velocities (middle row) and mean temperatures (bottom row) at 3 times of 0.007, 0.054 and 1.46 μs. Also shown are Cu and Ar mass fractions as well as mixture average quantities. Spatial scale is normalized to λ =10 μm.*

At the initial times after the Cu plume release, a small amount of highly energetic Cu atoms run away from the dense plume region and find themselves in a low-density background gas. The fact that the Cu atoms are heavier than the Ar atoms makes it even more difficult for them to dissipate energy in collisions with the background species and reach an equilibrium with the background gas. This mechanism is clearly evidenced in Figure 6 where we plot a typical instantaneous Cu atom velocity distribution function (VDF) are two locations: one inside the high-density Cu plume (explosion core) region and one in front of the propagating plume (at a distance of ~10–20 λ). The VDF inside the core is close to Maxwellian due to frequent collisions among Cu atoms. In front of the plume boundary (low-density region), a highly non-equilibrium VDF forms, which is characterized by two distant peaks in velocity space. One peak corresponds to the low-energy part of the VDF replicating that in the high-density core, and the other peak corresponds to the highly energetic Cu atoms escaped from the high-density plume zone and thus becoming almost collisionless. The VDF of such a form is characterized by high mean velocities and temperatures as observed in Figure 5 and Figure 6. The number density of the highly energetic atoms is almost 5 orders of magnitude lower than that in the Cu plume region as seen in Figure 6. This makes it difficult to accurately capture such a VDF with statistical DSMC methods making the DVM Boltzmann solver advantageous for modeling transient kinetic processes associated with the formation of energetic metal species during the metal plume expansion. The energetic metal atoms can propagate over long distances from the initial explosive core and can be detected experimentally. These neutral atoms can be easily ionized (e.g., in charge-exchange collisions) with surrounding plasma and become ions. The energy and angular distributions of ions can be a good diagnostic tool for the physical processes occurring in the small explosive zones far from detector locations.

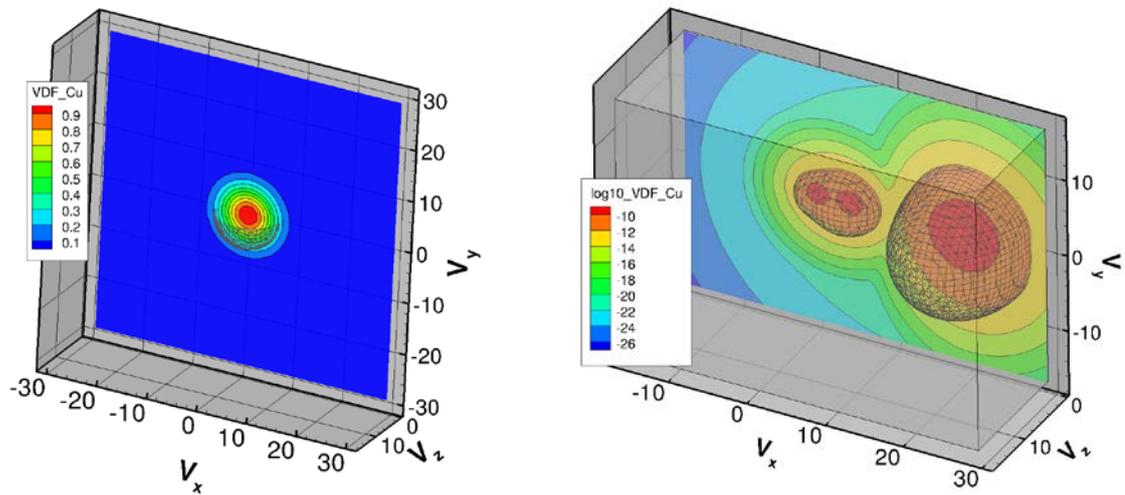

*Figure 6. Typical Cu atoms VDFs in 3D velocity space at two locations: explosion core (left, linear scale) and run-away zone (right, log scale). The velocities are normalized to $u_{ref}$ = 253 m/s.*

To study the influence of background gas on the dynamics of metal plume expansion, we have varied the background gas pressure. For the background gas pressures of 20 mTorr, the mean free path is $\lambda$ = 100 μm and for 2 Torr $\lambda$ = 1 μm. The results for the species temperatures for 20 mTorr are presented in Figure 7. One can see that for the lower background pressure, there is even stronger non-equilibrium, which manifests itself, among other things, in very high Cu atom temperatures reaching values of 120 times the background temperature. These high temperature regions extend over 100 mean free paths from the Cu propagation front. And because of the longer mean free path in the lower pressure conditions, the actual non-equilibrium zone is larger by a factor of at least several 10's than for 200 mTorr. The temporal evolutions of the species mean velocities and temperatures shown in Figure 8 indicate that the initial stages of the plume expansion are characterized by both high velocities (up to ~6500 m/s) and high temperatures (up to ~36,000 K).

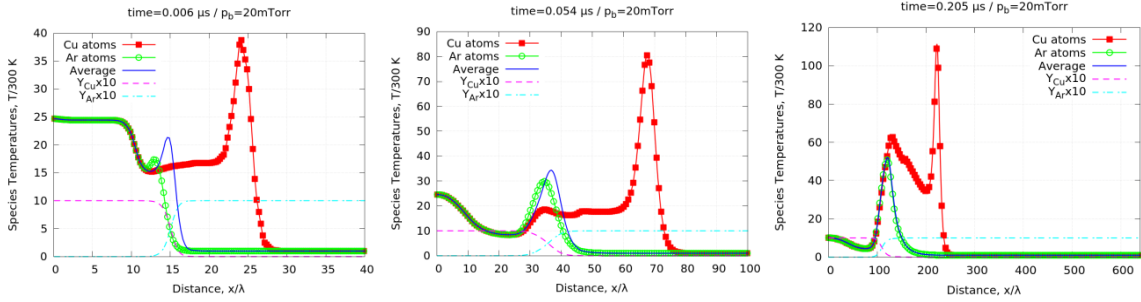

*Figure 7. Species macroparameters obtained by UFS-Boltzmann solver for 20 mTorr. Shown are species mean temperatures together with mixture average temperatures and species mass fractions at 3 times instances of 0.006, 0.054 and 0.2 μs. Spatial scale is normalized to λ =100 μm.*

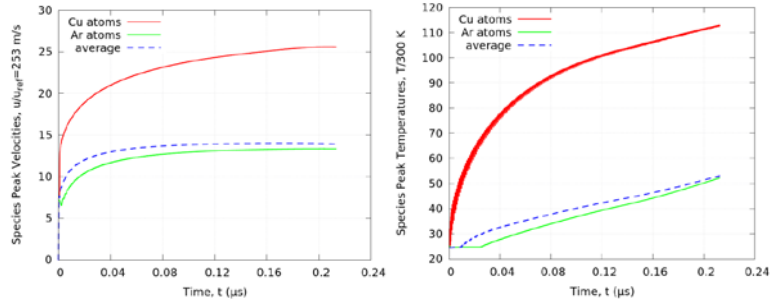

*Figure 8. Species peak mean velocities (left) and temperatures (right) as functions of time obtained by UFS-Boltzmann for 20 mTorr. Shown also are mixture average quantities.*

Splitting metal plume species into slow and fast components has been previously observed both experimentally and computationally using simplified models (see, e.g., Refs. [11,12]). It has been observed experimentally [12] that the plasma plume splitting occurs over time periods of up to several hundreds of ns and strongly influenced by the background gas pressure. Both these experimental observations are in excellent agreement with our predictions.

*3.3. 1D Simulations of Cu Plume Expansion into Argon with a Hybrid Solver*
The kinetic simulations of the high-density metal plume expansion into low-pressure Ar gas have clearly demonstrated that the initial stage of this process is highly non-equilibrium. The non-equilibrium zone extended up to 100 mean free paths in front of the metal expansion boundary. For simulation domains extending over several tens of thousands of mean free paths (see Figure 3), kinetic simulations are rather costly. Furthermore, the non-equilibrium zones typically exist over a short period of time during initial metal plume expansion when highly energetic metal atoms are generated. For later dynamics of the plume expansion, fluid models may be adequate.

Using UFS methodology, we have conducted hybrid kinetic-fluid simulations of the plume expansion into background gas. Details of the cell-by-cell selection of the kinetic and fluid solvers and coupling these solvers at kinetic-fluid interfaces have been previously described in [13,14]. A slightly modified continuum breakdown criterion based on the mean free path was used in present simulations, and the computational grid was adapted in the same way as in the previous simulations. By utilizing the dynamic load balancing capabilities for efficient partitioning of the kinetic and fluid cells, we obtained very good load balance across all computational cores utilized. The results for Ar pressure of 200 mTorr are presented in Figure 9. One can see that at the time instances shown, kinetic regions of about 100 and 200 mean free paths in length are required. By comparing with the full Boltzmann solutions, we could observe very good agreement, and a gain factor of up to 10 in terms of CPU time and required RAM have been achieved for a non-optimized kinetic breakdown criterion. Optimization of this criterion is expected to result in even more significant gain factors.

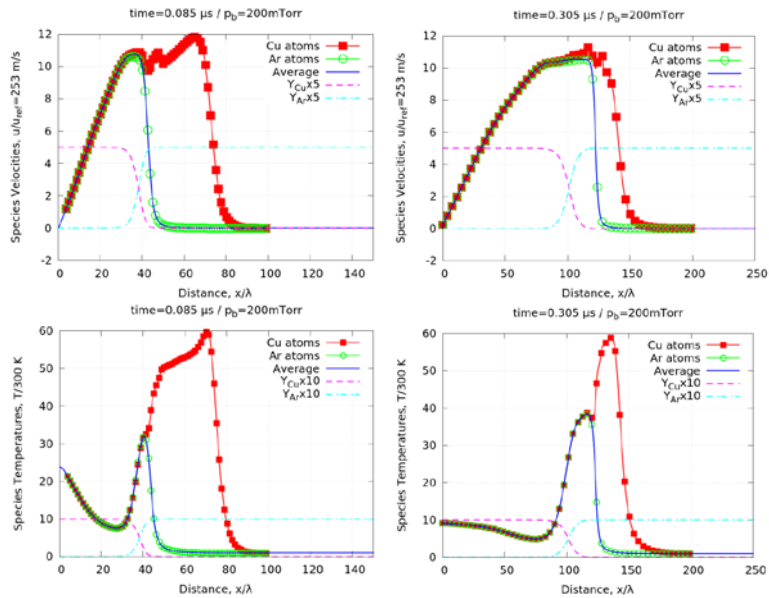

*Figure 9. Macroparameters obtained by hybrid solver for 200 mTorr: species mean velocities (top row) and mean temperatures (bottom row), at 3 times of 0.09 and 0.3 μs. Also shown are Cu and Ar mass fractions as well as mixture average quantities. Spatial scale is normalized to λ =10 μm.*

*3.4. 2D Axi-Symmetric Simulations of Cu Plume Expansion with the Navier-Stokes Solver*

We finally demonstrate the UFS capabilities for the problem of Cu plume expansion into a low-pressure Ar in multi-dimensional settings. We present here NS simulations of a spherical expansion of an initial high-pressure Cu core of 50 μm (5 Ar mean free paths) in radius with domain size of 80 $\lambda$. Snapshots of the Mach number and adapted grid are shown in Figure 10. We observe that due to the "snow-plow" mechanism, a supersonic shock wave propagates over background gas slowing down to subsonic velocities. Dynamicaly adapted grid allows resolving the large spatial scale disparity between the initial plume zone and the large expansion zone, as well as moving shocks and discontinuities.

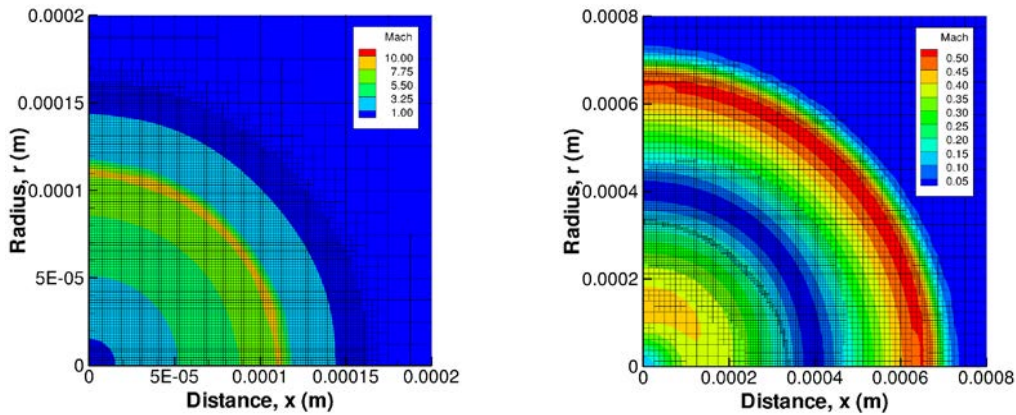

*Figure 10. 2D-axi NS solution for problem of metal plume free expansion for 200 mTorr. Shown are Mach numbers at t = 0.07 μs (left) and 0.7 μs (right) and dynamically adapted mesh.*

Furthermore, we have conducted simulations for metal atom injection over short periods of time in order to observe metal content propagation in 2D/3D setting. For these simulations, a high-pressure Cu source was set at the solid boundary in a shape of a cylinder (sphere in 3D). The opposite boundary is assumed to be a wall with zero velocity and room temperature boundary conditions. For a 25 ns source

(a short laser pulse), we have observed that the metal content exhibits strongly non-spherical, forward-directed pattern. Such patterns have been observed experimentally in many works (e.g., Ref. [12]). The dynamic AMR allows resolving the plume dynamics over a wide range of scales making the simulations very efficient. The initially strict CFL time step limitation was lifted for later stages and computations proceeded with larger time steps and muck coarser grids.

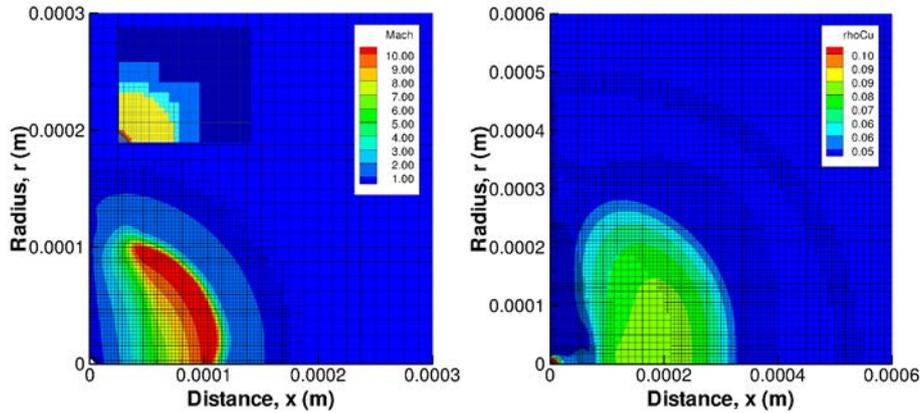

*Figure 11. 2D-axi NS solution for problem of metal plume expansion from a solid object for 200 mTorr. Shown are Mach number at t = 0.08 μs (left; insert shows close up view on initial grid with refinement levels from 4 to 9 around solid object) and Cu mass density at t = 0.52 μs (right). AMR grids at both time instances are also shown.*

### 4. Plasma Expansion into Vacuum

Plasma expansion into vacuum has been studied in a large number of works since publications by Gurevich et al. [15], due to high relevance to laboratory devices and applications as well as space science. Dense plasma expands into a vacuum due to thermal electron pressure; some ions can be accelerated to very high energies as a result of this process [16]. The electrons can be assumed to remain in equilibrium with the electric field since the expansion occurs on a time scale long compared with electron time scale [17].

In a series of recent works (see Ref. [18] and references cited therein), two regimes of plasma expansion have been distinguished. In the first regime, when electron motion is collisional, the process is largely similar to the conventional gas-dynamic expansion. In this regime, the electrons could remain Maxwellian during the expansion. In the second, collisionless regime, the assumption of Maxwellian electrons is not justified. A kinetic analysis of this problem in Ref. [18] has led to a conclusion that the deviation from the Maxwellian distribution fundamentally alters the ion acceleration with hot non-thermal electrons playing a critical role in the process [19].

*4.1. Implementation of fluid model for collisionless plasma expansion in UFS*

We have adapted UFS for simulations of collisionless plasma expansion. We assumed that electrons stay in equilibrium with the electrostatic potential, while the ion motion is described by two equations for density and mean velocity. The major difference between our model and other works is using the Euler framework. In a Lagrangian formulation,[21] the computational grid is moving with the ion front using analytically computed boundary conditions at the ion front. Such a formulation can provide more accurate results for 1d problems but is not practical for more general geometries (such as 2D and 3D). In our implementation, the electron density was updated at each time step using the Boltzmann relationship. We have implemented a Poisson-Boltzmann solver by extending the multi-grid Poisson solver in UFS. As a first step, we applied relaxation on electron density and potential during multi-grid and algebraic solver iterations. More sophisticated techniques can be used in future work, such as those involving non-linear Newton iterations with a defined Jacobian matrix, as well as coupled with electron and ion evolution strategies (see, e.g., Ref.[21]).

Results of 1D simulations of collisionless plasma expansion are shown in Figure 12. The electric field peaks at the locations of ion front decrease in magnitude as the ion front expands. A close up view on the ion front region is also shown in Figure 12 (bottom, right). One can see that the electric field has a pronounced peak at the ion front location, as expected for a plasma double layer where the ion density drops to low values within a short distance forming a space charge region. There are no peaks on the ion profiles at the plasma fronts, as in some predictions obtained by Lagrangian codes by other groups.[21] This can be explained by the fact that in our setup a finite computational domain (320 Debye lengths) was used and the initial ion profile was not an ideal step function due to the finite grid resolution. In Refs.[20,21], the issue of the initial ion front was discussed in detail: some works did find peaks of ion density, while others [22] did not find such peaks. The authors of Ref.[20] conclude that from an experimenter`s point of view, a finite width of the edge will always be present in laser-produced plasmas. We expect that the inclusion of collisions and multi-dimensional effects will relax the strict numerical constrains placed on solvers for modeling plasma expansion into vacuum.

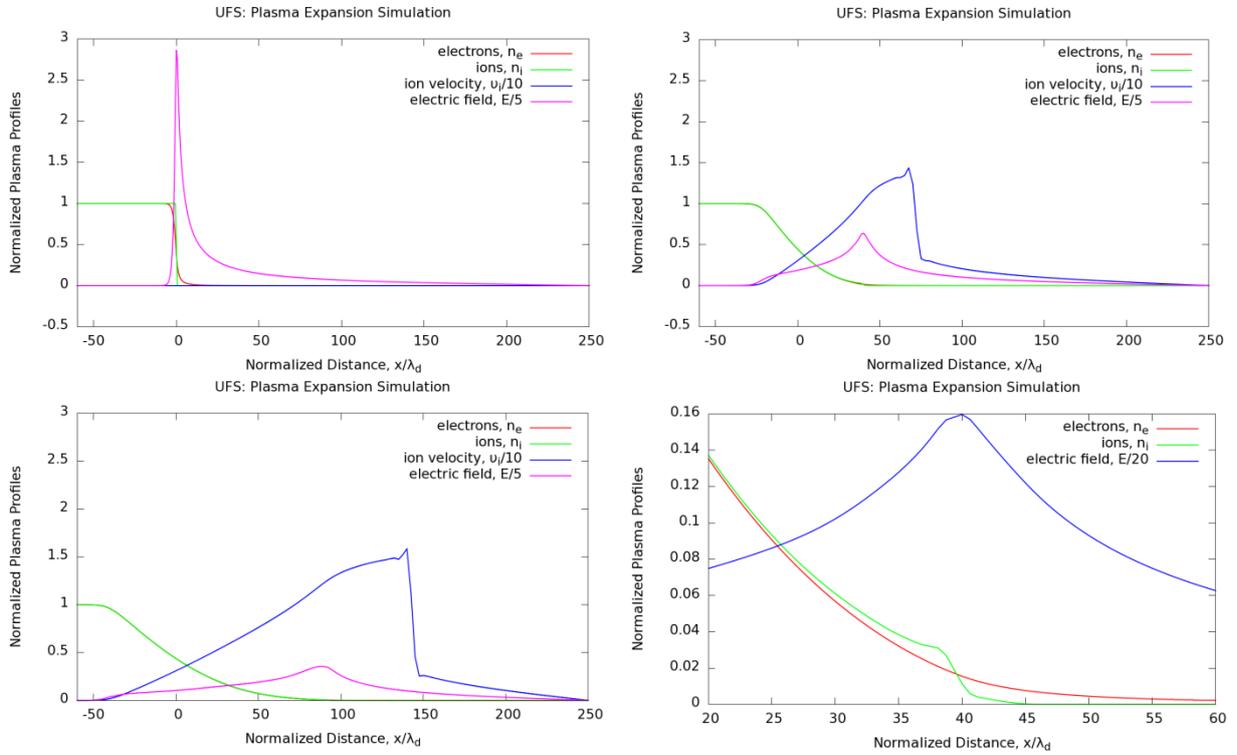

Figure 12. Electron and ion densities, ion velocity and electric field at normalized times 0 and 5 (top row) and 10 (bottom, left) for simulation of plasma expansion into vacuum. Bottom, right shows close up view of the ion front region at normalized time 5.

An important quantity to look at is the ion velocity at the location of the ion front. One can see in Figure 12 that the normalized velocities increase with time and reach values of around 1.5–2 at the normalized times between 5 and 10. These values, as well as the ion velocity time dependence, are consistent with those obtained in the literature; the predicted ion front velocities are slightly lower.

### 4.2. 3D Simulations of Collisionless Plasma Expansion

For 3D simulations of collisionless plasma expansion, the initial plasma density profile was assumed to be of a spherical shape with a radius of $20\,\lambda_d$. The Poisson-Boltzmann solver was used to compute self-consistent electron density and electrostatic potential distributions. The domain size was set to 240

$\lambda_d \times 240\, \lambda_d \times 240\, \lambda_d$. Grid adaptation strategy was based on gradients of electron and ion densities, and the space charge. Time evolution of ion density is shown in Figure 13 at several time instances. One can see that at early stages plasma expansion (t < 1) is close to spherical. However, at t > 1, a flat-shaped ion front forms, which moves with a slowly increasing speed until t < 3–4. For times t > 5, the flat-shaped ion front starts to collapse inward while undergoing self-focusing and forming a well-defined bunch at later times. The dynamically adapting grid allows capturing lower density ion bunch at later times when its amplitude drops by a factor of 20–30. Resolving the low-density ion density at long distances from the explosion is crucial for successful comparison with the experimentally measured ion distributions.

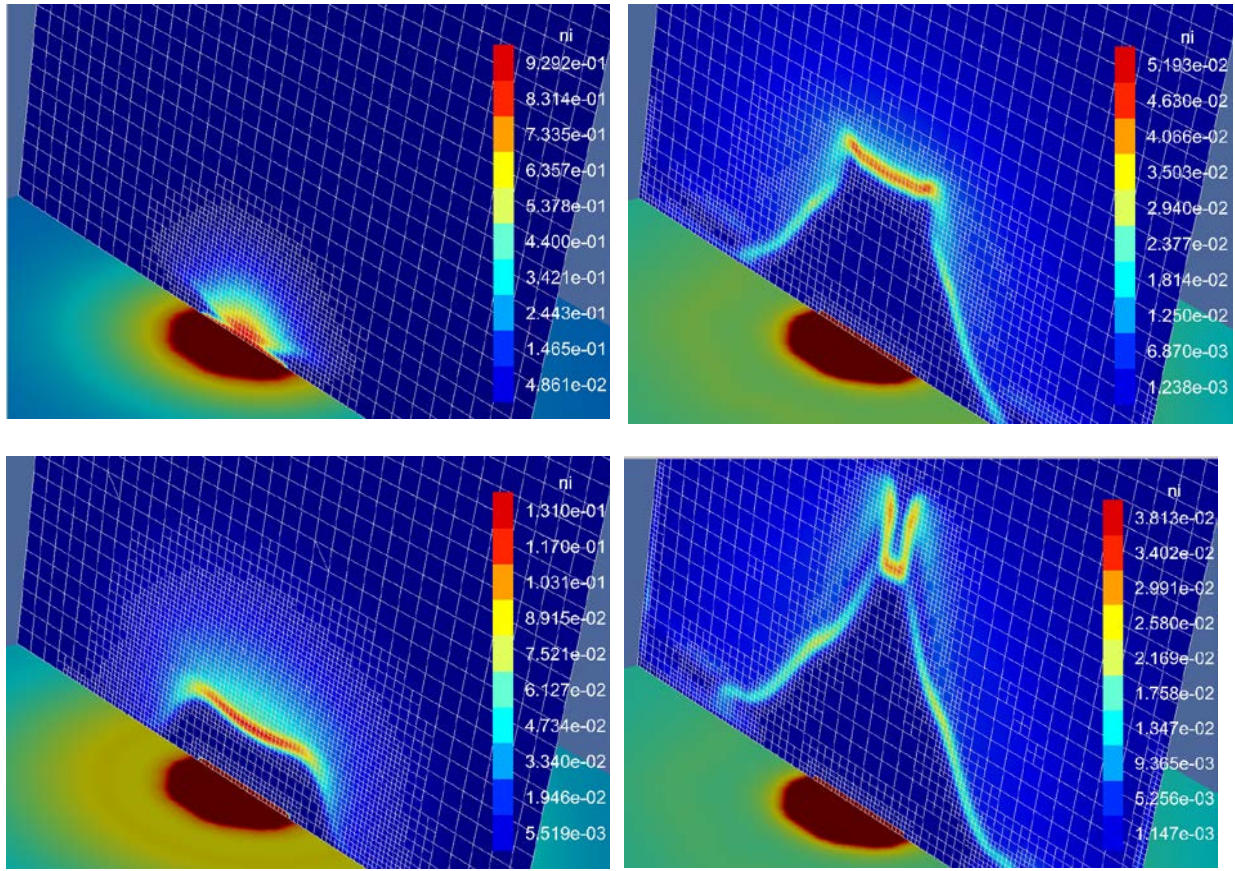

*Figure 13. Contours of ion density at four time instances: 1, 3 (left column, top to bottom), 5 and 6 (right column, top to bottom), together with adapted grid for 3D simulation of plasma expansion into vacuum.*

## 5. Conclusions

We have investigated gas dynamic and plasma effects for expanding plasma jets. The Unified Flow Solver (UFS) has been adapted and enhanced for simulations of mixed continuum-rarefied flows, transient gas plume dynamics and collisionless plasma expansion. The enhancements allowed us to efficiently reconstruct a 3D picture of plasma expansion into vacuum. We have observed an interesting phenomenon of ion bunching in our 3D simulations. Since a simplified model of plasma expansion has been used, more accurate predictions and comparison with experiments will require implementations of additional physics. We have demonstrated that the UFS framework is capable simulating different aspects of plasma expansion problems. We identified further advances that are required for hybrid fluid-kinetic simulations of laser-ablation and EEE-induced plasma jets.

**Acknowledgments**

This work was partially supported by the DOE SBIR Project DE-SC0015746, by the US Department of Energy Office of Fusion Energy Science Contract DE-SC0001939, and by the NSF EPSCoR project OIA-1655280. Thanks are due to Anna Frolova for help with the code development.